\documentclass[showpacs,preprint,showkeys]{revtex4-1} 
\usepackage{amsfonts,amsmath,amssymb}
\usepackage{enumerate}
\usepackage{hyperref}
\usepackage{graphicx}
\newcommand{\be}{\begin{equation}}
\newcommand{\ee}{\end{equation}}
\newcommand{\bea}{\begin{eqnarray}}
\newcommand{\eea}{\end{eqnarray}}
\newcommand{\bes}{\begin{subequations}}
\newcommand{\ees}{\end{subequations}}

\newcommand{\cN}{{\cal N}}

\newcommand{\Tr}{\mbox{Tr}}

\begin{document}
\setcounter{page}{0}
\title{Probing Non-Toric Geometry with Rotating Membranes}
\date{\today}
\author{Jung Hun Lee}
\author{Sunchang Kim}
\author{Jongwook Kim}
\author{Nakwoo Kim}
\email{nkim@khu.ac.kr}
\affiliation{Department of Physics and Research Institute of Basic Science, \\
Kyung Hee University, Seoul 130-701, Korea}
\begin{abstract}
Recently Martelli and Sparks presented the first non-toric 
$\mbox{AdS}_4/\mbox{CFT}_3$ duality relation between 
M-theory on 
$\mbox{AdS}_4\times V_{5,2}/\mathbb{Z}_k$ and 
a class of three-dimensional
$\cN=2$ quiver Chern-Simons-matter theories. $V_{5,2}$ is a seven-dimensional homogeneneous Sasaki-Einstein manifold with isometry group
$ SO(5)\times U(1)_R$, which is in general
broken to $SU(2)\times U(1)\times U(1)_R$ by the orbifold projection if $k>1$. 
The dual field theory is 
described by the ${\cal A}_1$ quiver, $U(N)_k\times U(N)_{-k}$ gauge group, four bifundamentals, two
 adjoint chiral multiplets interacting via a cubic superpotential. 
We explore this proposal by studying various classical membrane solutions moving
in $V_{5,2}$. 
Rotating membrane solutions of 
folded, wrapped, spike, and giant magnon types are presented with their dispersion
relations. We also discuss their dual operators in the Chern-Simons-matter theory.
\end{abstract}
\maketitle
\thispagestyle{empty}

\pagestyle{plain}
\section{Introduction}
Over the last two years there has been remarkable progress on the understanding
of M2-brane dynamics. It is now widely accepted
 that multiple M2-branes can be described by Chern-Simons-matter theories. More concretely, 
the most well-established relation dictates M2-branes at
orbifold singularity $\mathbb{C}^4/\mathbb{Z}_k$ are described by
 $\cN=6$ supersymmetric
quiver Chern-Simons-matter theory with 
$U(N)_k\times U(N)_{-k}$ gauge symmetry and four bifundamental chiral multiplets \cite{Aharony:2008ug}.
 The order of orbifold group $k$ appears
as the quantized Chern-Simons level in the field theory. In terms of 
AdS/CFT, the dual is M-theory in $\mbox{AdS}_4\times S^7/\mathbb{Z}_k$
background.

It is certainly an interesting problem to find new $\mbox{AdS}_4/\mbox{CFT}_3$
duals with less supersymmetry. We now have a large number of such proposals, see 
for instance \cite{
Imamura:2008nn,Martelli:2008si,Hanany:2008cd,Imamura:2008qs,
Hanany:2008fj,Franco:2008um,Franco:2009sp,Aganagic:2009zk,
Martelli:2009ga}.
One may utilize the brane construction technique \cite{Elitzur:1997fh}
and write Chern-Simons duals for 
various orbifolds of $\mathbb{C}^4$. Or one turns to 
the brane tiling method \cite{Franco:2005sm,
Davey:2009et} if the ambient geometry is toric.

In general we expect a ${\cN=2}$ superconformal field 
theory for a Freund-Rubin type background of $\mbox{AdS}_4\times {\cal M}_7$,
if ${\cal M}_7$ is a seven dimensional Sasaki-Einstein manifold. In canonical
form a Sasaki-Einstein manifold is written as a twisted $U(1)$ fibration over 
a K\"ahler-Einstein manifold. $\mathbb{CP}^3$ leads to the trivial example of
$S^7$, while $\mathbb{CP}^1\times\mathbb{CP}^1\times\mathbb{CP}^1$ and 
$\mathbb{CP}^2\times\mathbb{CP}^1$ lead to so-called $Q^{1,1,1}$ and $M^{1,1,1}$ 
respectively. Since the rank of isometry group is four, above are all examples of 
{\it toric} manifolds. The dual field theories are all given as
quiver Chern-Simons models.

In order to establish the duality, one first computes the vacuum moduli space
of the field theory. In general one recovers a discrete quotient, or an orbifold
of the desired Sasaki-Einstein space. Then we have to check if the spectrum 
of chiral primary operators is consistent with Kaluza-Klein spectrum of the
11 dimensional supergravity and orbifolds thereof.
It turns out to be
 crucial to include monopole operators in order to see the symmetry 
enhancement for $k=1$. For more details on $Q^{1,1,1}$ and $M^{1,1,1}$, 
see \cite{Martelli:2008si} and \cite{Franco:2009sp}.

One can go beyond the particle limit and consider membrane dynamics directly. 
Although it is not known how to quantize membranes in a nontrivial
background, one can use classical solutions with large energy as an approximation
of the full quantum result. Such a semiclassical approach of AdS/CFT is initiated 
in \cite{Gubser:2002tv}, and for the applications in the 
context of M2-branes see for instance \cite{Bozhilov:2005ew,*Bozhilov:2005xs,
*Bozhilov:2006gh,*Bozhilov:2007km,*Bozhilov:2007mb,*Bozhilov:2007bi,
*Bozhilov:2007wn,*Wen:2007az,*Ahn:2008gd,Kim:2010ck}. One can study explicit
rotating membrane solutions, which are dual to gauge invariant
composite operators with large conformal dimension. The dispersion relations
between energy and angular momenta are believed to provide nontrivial 
quantitative prediction on the conformal field theory.

We continue our previous work \cite{Kim:2010ck} and study rotating membranes
in $\mbox{AdS}_4 \times {V}_{5,2}$ in this paper. 
$V_{5,2}$ is another example of seven dimensional
Sasaki-Einstein manifold. It is a coset $SO(5)\times SO(2)/SO(3)\times SO(2)$ 
so homogeneous, but non-toric since its isometry is $U(1)\times SO(5)$ and rank 3.
The dual field theory is given as a relatively simple quiver Chern-Simons theory
with cubic superpotential \cite{Martelli:2009ga}. The spectra of chiral 
primary operators are consistent with the Kaluza-Klein analysis of 11 dimensional
supergravity on $V_{5,2}$ reported in \cite{Ceresole:1999zg}.

We probe this non-toric duality using membranes. Explicit classical solutions
are constructed and we provide the dispersion relations. As we wrap the membranes
along the great circle of an $S^2$ within $V_{5,2}$, the membrane dynamics is 
effectively reduced to that of Polyakov strings. We study configurations 
analogous to folded, wrapped strings as well as giant magnons \cite{Hofman:2006xt} 
and single spikes \cite{Kruczenski:2004wg}. We will discuss what dual operators 
would look like.

This paper is organized as follows. Sec.\ref{2} will be a review of \cite{Martelli:2009ga}
and we present both 
the supergravity background and quiver Chern-Simons theories, mainly to  
setup the notations. Sec.\ref{3} is the main part where we report on 
the membrane solutions and discuss the field theory interpretations.
We will conclude in Sec.\ref{4} with discussions.

\section{\label{2}$\mbox{AdS}_4 \times {V}_{5,2}$ and the dual Chern-Simons theory}
\subsection{The supergravity background}
Let us start by presenting the M-theory background of our interest. 
We will closely follow \cite{Martelli:2009ga}, although
occasionally we employ different conventions for later conveniece. For the 
geometry of $V_{5,2}$, see also \cite{Bergman:2001qi}.

The 11 dimensional
 metric is given as a direct product of $\mbox{AdS}_4$ and a seven-dimensional
space $V_{5,2}$.
\bea
\mbox{d}s^2_{11} &=& L^2(
\tfrac{1}{4}\mbox{d}s^2_{\text{AdS}_4} + \mbox{d}s^2_{V_{5,2}}), \\
\mbox{d}s^2_{\text{AdS}_4} &=& -\cosh^2\rho\, \mbox{d}t^2 + \mbox{d}\rho^2 +\sinh^2\rho\,(\mbox{d}\vartheta^2 + \sin^2\theta \mbox{d}\varphi^2), \\
\mbox{d}s^2_{V_{5,2}} &=& \tfrac{9}{64}\left[\mbox{d}\psi + \cos\alpha(\mbox{d}\beta + \cos\theta_1 \mbox{d}\phi_1 + \cos\theta_2 \mbox{d}\phi_2)\right]^2
+ \mbox{d}s^2_{Gr_{5,2}}.
\eea
In the above the Sasaki-Einstein space $V_{5,2}$ is expressed in canonical form, i.e.
$U(1)$-fibration over K\"ahler-Einstein space. $G_{5,2}=SO(5)/SO(3)\times SO(2)$ is a coset space, 
so it is homogeneous
and the metric can be given as 
\bea
     \mbox{d}s^2_{Gr_{5,2}}&=&  \tfrac{3}{32}\left[
4\mbox{d}\alpha^2 + \sin^2\alpha(\mbox{d}\beta + \cos\theta_1 \mbox{d}\phi_1 + \cos\theta_2 \mbox{d}\phi_2)^2 \right. \nonumber\\
          && + (1 + \cos^2\alpha)(\mbox{d}\theta^2_1 + \sin^2\theta_1 \mbox{d}\phi^2_1 + \mbox{d}\theta^2_2 + \sin^2\theta_2 \mbox{d}\phi^2_2) \nonumber\\
          && + 2\sin^2\alpha\cos\beta\sin\theta_1\sin\theta_2 d\phi_1 \mbox{d}\phi_2 - 2\sin^2\alpha\cos\beta \mbox{d}\theta_1 \mbox{d}\theta_2 \nonumber\\
          && \left. - 2\sin^2\alpha\sin\beta(\sin\theta_2 \mbox{d}\phi_2 \mbox{d}\theta_1 + \sin\theta_1 \mbox{d}\phi_1 \mbox{d}\theta_2)
\right] . 
\eea
The ranges of the coordinates are
\bea
0\leq\theta_1,\theta_2\leq\pi, \quad 0\leq\phi_1,\phi_2 < 2\pi, \quad 0\leq\psi <4\pi, \quad 0\leq\alpha\leq\frac{\pi}{2}, \quad 0\leq\beta <4\pi .
\eea
One can compute the volume of $V_{5,2}$ and obtain 
\bea
\text{Vol}(V_{5,2}) = \frac{27}{128}\pi^4\,\, .  \label{vol}
\eea

It will be particularly important for us that 
and the isometry group of $V_{5,2}$ 
is $SO(5)\times U(1)_R$. Since the rank of the group is three while
the complex dimension of the cone over $V_{5,2}$ is four, we have a {\it non-toric} manifold.

The curvature of the above metric is sourced by a four-form flux,
\bea
G^{(4)}    &=& \frac{3L^3}{8}\rm{Vol}({\text{AdS}_4})\,.
\eea
The quantization of the G-flux allows us to relate $L$ with the number of M2-branes, i.e.
\bea
L^6 = \frac{(4\pi l_p)^6 N}{81\pi^4}\, , 
\eea
where $l_p$ is the eleven-dimensional Planck length. 
\subsection{The quiver Chern-Simons-matter theory}
\begin{figure}
\includegraphics[scale=0.45,viewport = 200 150 300 350]{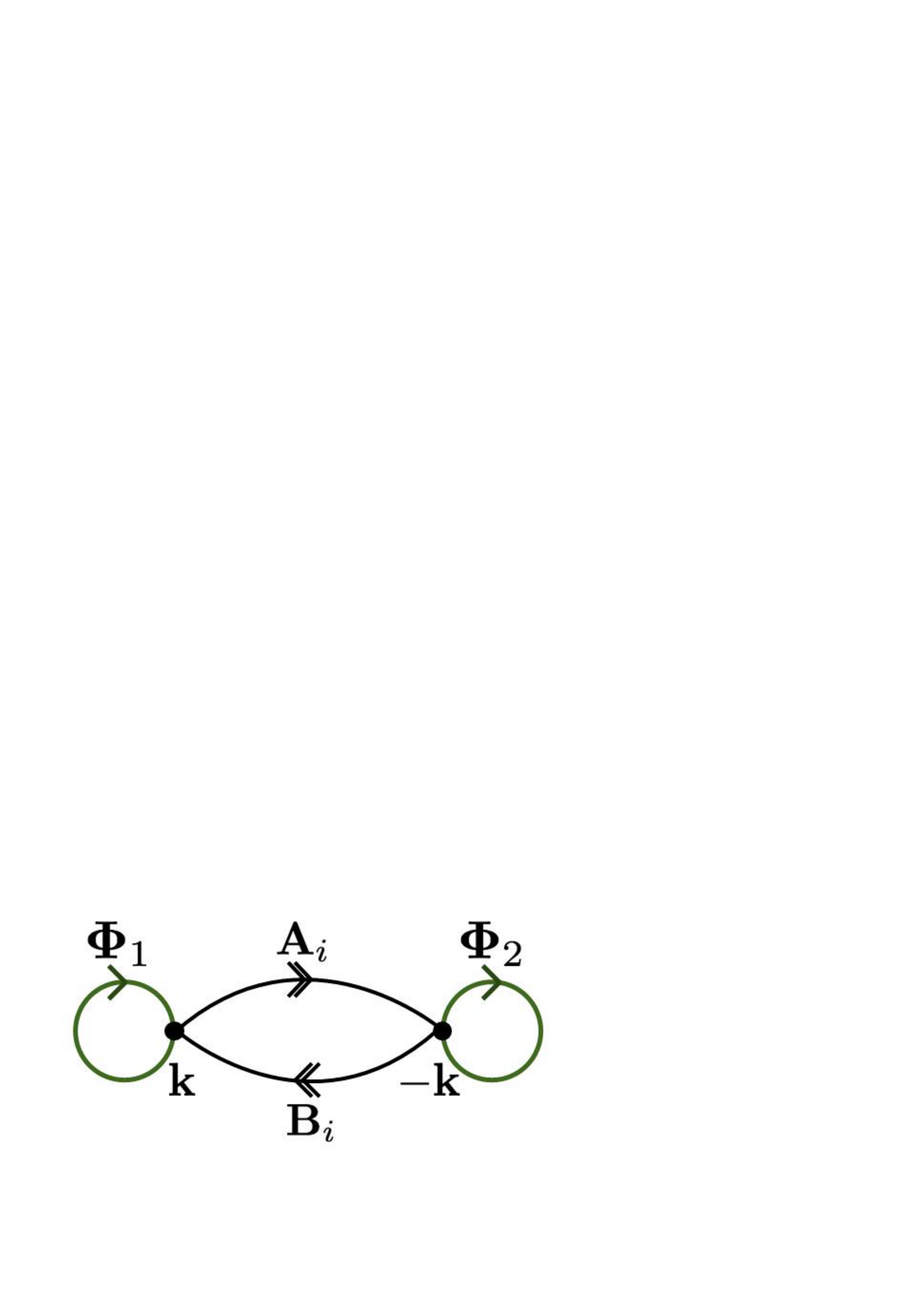}
\caption{\label{fig}The quiver diagram dual to $\mbox{AdS}4\times V_{5,2}$.}
\end{figure}
Now we move to the dual conformal field theory side. According to 
\cite{Martelli:2009ga}, the Chern-Simons-matter
 theory described by ${\cal A}_1$ quiver in FIG.~\ref{fig} is dual to 
M-theory in $\mbox{AdS}_4\times V_{5,2}/\mathbb{Z}_k$.
 The gauge symmetry of the conformal field theory
is $U(N)\times U(N)$, with Chern-Simons levels
$(k,-k)$. As it is obvious in the quiver diagram, $\Phi_1$ and $\Phi_2$ are in adjoint
representation of each $U(N)$. The matter fields $A_1,A_2$ are in $(N,\bar{N})$, while
$B_1,B_2$ are in $(\bar{N},N)$ representations. In total, there are thus
$6N^2$ chiral multiplets
of $\cN=2$ supersymmetry in three dimensions. 

In addition to the gauge couplings, the interactions are described by the following
cubic superpotential.
\bea
W = \Tr[{s}(\Phi^{3}_1 + \Phi^{3}_2) + \Phi_1 (A_1B_2 - A_2B_1)+ 
\Phi_2(B_2A_1 - B_1A_2)]\, , 
\eea
where $s$ is a complex-valued coupling constant. Obviously 
all the elementary fields should 
be given R-charges $2/3$. 
One can easily check that $W$ is invariant under $SU(2)$
if $(A_1,A_2)$ and
$(B_1,B_2)$ are both doublets. One discovers there is another $U(1)$, with charge
assignments $0,1/2,-1/2$ for $\Phi_i,A_i,B_i$ respectively. As we will see later, for $k=1$
the non-R symmetry $SU(2)\times U(1)$ is enhanced to $SO(5)$.

For abelian case $N=1$, one can 
easily see that the F-term conditions imply 
\be
\Phi_1+\Phi_2=0 , \quad 3s \Phi_1^2 + A_1B_2 - A_2 B_1 = 0 .
\label{fterm}
\ee
If one defines the new variables as 
 $z_1={1\over 2}(A_1 + B_2)\,, z_2={i\over 2}(A_1 - B_2)\,, z_3={i\over 2}(A_2 + B_1)\,,
z_4={1\over 2}(A_2 - B_1)\,$ and $z_0 = (3s)^{1/2}\Phi_1$, we have an equation
\bea
X_2 \equiv \{z_0^2 + z_1^2 + z_2^2 + z_3^2 + z_4^2 = 0\}\,,
\label{a1}
\eea
for the vacuum moduli space. 

Obviously $X_2$ is a complex four dimensional space with $A_1$-type singularity
at the origin. Like a conifold which is the three dimensional version \cite{Candelas:1989js},
 $X_2$ is Calabi-Yau and one can assign a K\"ahler and Ricci-flat metric. $X_2$ can be also 
viewed as a metric cone, and it turns out that 
$V_{5,2}$ is the base manifold. It is thus conjectured \cite{Martelli:2009ga} that 
the quiver Chern-Simons theory in FIG.~\ref{fig} has an IR fixed point which 
is dual to the near horizon limit of M2-branes 
of M2-branes at the hypersurface singularity of $X_2$.

In principle one has to identify the points along the gauge orbits, but the baryonic
$U(1)$ gauge transformation is used to fix the phase of the dual photon field. 
Of course this is  the same as what happens in the ABJM model \cite{Aharony:2008ug}.
For generic values of $k>1$, the gauge invariance requires we identify
\be
(A_1, A_2, B_1, B_2 ) \sim 
( e^{\frac{2\pi i}{k}} A_1,e^{\frac{2\pi i}{k}} A_2, 
e^{-\frac{2\pi i}{k}}B_1, e^{-\frac{2\pi i}{k}} B_2 ) . 
\label{quot}
\ee
The flavor symmetry $SU(2)_A\times SU(2)_B$ for doublets $A_i$ and $B_i$, 
although not manifest in the field theory, is translated to the 
two two-spheres parametrized by $\theta_i,\phi_i$ in $V_{5,2}$. The quotient in 
Eq.~(\ref{quot}) should correspond to 
\be
(\phi_1, \phi_2) \sim (\phi_1+\tfrac{2\pi}{k},\phi_2-\tfrac{2\pi}{k})\, .
\ee
It is then obvious that  the isometry $SO(5)$ is broken to $SU(2)\times U(1)
\subset SU(2)\times SU(2) \subset SO(4) \subset SO(5)$, for $k>1$.

We are now ready to consider chiral primary operators and see how they 
are arranged in terms of $U(1)_R \times SU(2) \times U(1)$. Let us first consider
abelian $N=1$ case with large $k$. As usual, we enumerate gauge singlet operators up to 
F-term conditions. At order one, we have only one neutral operator, $\Phi_1$ (or $-\Phi_2$).
This operator carries both conformal dimension and R-charge $2/3$, but it is a singlet and neutral
for $SU(2)\times U(1)$. At order two, after taking Eq.~(\ref{fterm}) into
account, we have four independent operators: 
$A_1B_1,A_2B_2,\tfrac{1}{\sqrt{2}}(A_1B_2+A_2B_1)$, and 
$\Phi_1^2$. The first three obviously constitute a triplet of $SU(2)$ and has $U(1)$ 
charge
$1$. $\Phi_1^2$ is invariant under $SU(2)\times U(1)$. All of them have both 
conformal dimension and R-charge $4/3$. One can easily convince oneself that this 
pattern will persist with higher order operators. At level $n$, the operators have 
conformal dimension and R-charge $2n/3$. The $SU(2)$ spin and $U(1)$ charge should take 
values between $0$ and $n/2$.

It is the monopole operators which play a crucial role in the symmetry enhancement 
for small values of $k$. The mechanism is again very similar to the ABJM model \cite{Aharony:2008ug}
and other examples of duality for three dimensional Chern-Simons field theories.
See for instance \cite{Martelli:2008si,Franco:2009sp,Kim:2010ck} for homogeneous Sasaki-Einstein manifolds
$Q^{1,1,1}$ and $M^{1,1,1}$. A monopole operator for abelian case
is the dual photon field $e^{ia}$
which carries charges $(k,-k)$ for gauge fields $U(1)\times U(1)$. For $k=1$, we can
essentially make every combination $A^{m_1}B^{m_2}$ neutral, if we allow insertions of an 
arbitrary number of monopole operators. Put differently, we can use the neutral combinations $A_i e^{-ia},B_i e^{ia}$ 
as the new {\it alphabets} in constructing {\it words} of chiral primary operators. Under
the assumption that the monopole operators do not change the conformal dimension, one
obtains $SO(5)$ since 
there is now only the F-term condition Eq.~(\ref{a1}) to consider. 

In general for nonabelian theories with $k=1$, one can express the chiral primary operators very schematically as 
\be
\Tr [\Phi^{n_1}(AB)^{n_2}(Ae^{-ia})^{m_1 } (Be^{ia})^{m_2}]\, . 
\ee
For the comparison with rotating membranes, it is useful to record the possible
$SU(2)_A\times SU(2)_B$ representations in addition to the R-charge 
\be
R = {2\over 3}[n_1 + 2 n_2 + m_1 + m_2] . 
\ee
Since the number of $A$'s is $n_2+m_1$, the spin for $SU(2)_A$ ranges up to 
$(n_2+m_1)/2$. For $SU(2)_B$, the maximum spin is obviously $(n_2+m_2)/2$.

\section{\label{3}Rotating membranes in $V_{5,2}$}
As in our previous work on $\mbox{AdS}_4\times M^{1,1,1}$ \cite{Kim:2010ck}, 
for the membrane dynamics
we will use the partly gauged-fixed version of Polyakov-type action which is devised by 
Bozhilov in \cite{Bozhilov:2005ew}.
\be
S = \int \frac{\mbox{d}^3\sigma}{4\lambda^0}\left[G_{00}-(2\lambda^0T_2)^2 \det G_{ij}\right]\, ,\label{action}
\ee
where $G_{mn} = \partial_m X^M\partial_n X^Ng_{MN}(X)$ is the
induced metric on the membrane worldvolume.
$T_2$ is the tension and $\lambda^0$ is a Lagrange multiplier. In addition to the
Euler-Lagrange equations derived from Eq.(\ref{action}), we have constraint equations
\bea
G_{00} + (2\lambda^0T_2)^2\det G_{ij} &=& 0\label{constraint1}\, , \\
G_{0i} &=& 0\label{constraint2}\, .
\eea
With some wisely chosen ans\"atze, the equations of motion 
and constraints are either trivially satisfied
or derivable from an auxiliary mechanical system. In this section
 we study various nontrivial
solutions of folded/wrapped rotating membranes, and also giant magnons and spinning spikes.

Since we are only interested in the membrane motion in $\mathbb{R}_t \times V_{5,2}$, 
we will set the coordinates in $\mbox{AdS}_4$ to
constants, i.e. $\rho = \vartheta = \varphi = 0$. 
Furthermore, we choose the temporal gauge and set $t=\kappa \tau$. We will consider
 membranes
spinning along $\psi,\beta,\phi_1,\phi_2$ angles. Once we find explicit solutions, 
we  compute the angular 
momenta conjugate to $\psi,\phi_1,\phi_2$ and identify them as the R-charge, spins for 
$SU(2)_A$ and $SU(2)_B$ respectively. 
\subsection{\label{3-1}Particle-like solutions and chiral primaries
}
In this section we consider membranes collapsed to a point. In other words, we will 
simply study the geodesic motion and evaluate the conserved quantities. It turns out
that without losing generality we can set $\psi=\omega\tau$ and all other angles to 
arbitrary constants. One can easily check that all the equations of motion are satisfied
as well as the constraints, provided $\kappa^2=\tfrac{9}{16}\omega^2$. 

The conserved charges are given as 
\be
Q_t = \tfrac{\sqrt{\lambda'}}{4}\kappa,\quad
Q_{\psi} = \tfrac{9\sqrt{\lambda'}}{64} \omega ,\quad
Q_{\phi_1} =\tfrac{9\sqrt{\lambda'}}{64}\omega\cos\alpha\cos\theta_1, \quad
Q_{\phi_2} = \tfrac{9\sqrt{\lambda'}}{64}\omega\cos\alpha\cos\theta_2  \, , 
\ee
with $\sqrt{\lambda'} = L^2/2\lambda^0$.

It is natural to identify $E\equiv Q_t$ as the conformal dimension $\Delta$ 
of the dual field
theory operator. 
Then one can see       that 
\be
E = \tfrac{4}{3} |Q_{\psi}| \ge \tfrac{4}{3} |Q_{\phi_i}| , 
\ee
which can be compared to the field theory side relation for chiral primary operators,
\be
\Delta = |R| \ge \tfrac{8}{3}|J_A|, \tfrac{8}{3}|J_B| . 
\ee
We thus choose to identify $R=\tfrac{4}{3}Q_{\psi},
J_A=\tfrac{1}{2}Q_{\phi_1},J_B=\tfrac{1}{2}Q_{\phi_2}$ from now on. 

It is easy to write down the dual operators for the particle solutions
with different constant values of $\alpha,\theta_1,\theta_2$. If $\kappa=3\omega/4>0$,
the dual operators are holomorphic expressions of $\Phi,A_i,B_i$ and $E=R$. 
For $\alpha=\pi/2$, $J_A=J_B=0$ and the dual operators should be $\Tr \,\Phi^{3R/2}$.
Another extreme example is at $\alpha=\theta_1=\theta_2=0$, which leads to $\Delta=R=\tfrac{8}{3}J_A=\tfrac{8}{3}J_B$. They are dual to $\Tr (A_1B_1)^{3R/4}$.
\subsection{\label{3-2}Simple classes of rotating wrapped membranes}
From this point on, we will consider membrane solutions occupying a nontrivial
two dimensional space at every instance. 
Before we move to nontrivial folded/wrapped type solutions in the next subsection, 
we present here
relatively simple solutions with a quadratic dispersion relation. 

Consider  
the following configuration. 
\be
\psi=\omega\tau, \quad \alpha=n\sigma_1, \quad \theta_1=\theta_2=m \sigma_2, \quad
\beta = \pi \mbox{ or } 3\pi , 
\ee
and $\phi_1,\phi_2$ are set to arbitrary constants. The winding numbers $n,m$ are integers.
All the equations are either trivially satisfied or equivalent to 
\be
\kappa^2 = \tfrac{9}{16}\omega^2 + (2\lambda^0 T_2 L)^2 \tfrac{9n^2m^2}{16}. 
\ee
It is easy to compute the conserved quantities, 
\bea
E = \tfrac{\sqrt{\lambda'}}{4}\kappa ,\quad
R = \tfrac{3\sqrt{\lambda'}}{16}\omega , \quad
J_A = J_B = 0 . 
\eea
Combining these two equations, we have the dispersion relation
\be
E^2 = R^2 + \left(\tfrac{3nm\sqrt{\lambda}}{16}\right)^2 . 
\ee
Note that the gauge parameter $\lambda^0$ disappers in the dispersion relation. We have defined $\sqrt{\lambda}=T_2 L^3$, which corresponds to 't Hooft coupling constant 
on the dual field theory side. 

One can also check that configurations given as 
\be
\psi=\omega\tau,  \quad \theta_1=\theta_2=m \sigma_2, \quad
\beta = 0 \mbox{ or } 2\pi , 
\ee
with
\be
\quad \sin\alpha=2n\sigma_1/\pi 
\label{meri}
\ee
also satisfies the equations, if 
\be
\kappa^2 = \tfrac{9}{16}\omega^2 + (2\lambda^0 T_2 L)^2 \tfrac{9n^2m^2}{4\pi^2}. 
\ee
Here $\phi_1,\phi_2$ are again constants, and the parameters 
$n,m$ are integers. Note that Eq.~(\ref{meri})
describes a circle wrapping $n$-times along a {\it meridian} of $V_{5,2}$, 
if all other coordinates were constants.
Strictly speaking Eq.~(\ref{meri}) makes sense only if $0\le \sigma_1 \le \pi/2n$, 
since $0\le \alpha \le \pi/2$. We can however continuously extend the geodesic up to 
$\sigma_2=2\pi$ as we do with $\theta$, while we traverse along the great circle $n$-times.

For conserved quantities, we again find $
J_A = J_B = 0
$, and a dispersion relation
\be
E^2 = R^2 + \left(\tfrac{3nm\sqrt{\lambda}}{8\pi}\right)^2 . 
\ee

Let us now contemplate on the form of dual operators. First of all, we argue that they must
be near-BPS and written as a  holomorphic expression for $R>0$ because the dispersion relation 
approaches the unitarity bound $E\ge R$ as $E,R\rightarrow \infty$. 
Note that
 we have effectively identified the two $S^2$'s and the solution is invariant under the exchange of two 
$SU(2)$'s. We thus expect there are no monopole operators and we can 
write schematically the dual operators as $\Tr [\Phi^{n_1} (AB)^{n_2}]$, with $n_1+2n_2=3R/2$.
Also recall that among four possible combinations of $A_iB_j \, (i,j=1,2)$, the $SU(2)$ singlet
is automatically projected out due to the F-term condition. The fact that we are left with {\it
symmetric} combinations is in harmony with $\theta_1=\theta_2$.

\subsection{Rotating membranes in the subspace $T^{1,1}$}
Let us now consider multi-spin rotating membranes. We will restrict ourselves to 
the subspace defined by $\alpha=\beta=0$. One can easily check that this is a consistent 
truncation with all equations. Effectively we are led to study membranes in 
$\mathbb{R}_t\times T^{1,1}$, with the following metric. 
\bea
\tfrac{1}{L^2}\mbox{d}s^2 &=& -\tfrac{1}{4}\mbox{d}t^2 + \tfrac{9}{64}\left(\mbox{d}\psi + \cos\theta_1 \mbox{d}\phi_1 + \cos\theta_2 \mbox{d}\phi_2\right)^2 
\nonumber \\  && 
+\tfrac{3}{16}(\mbox{d}\theta^2_1 + \sin^2\theta_1 \mbox{d}\phi^2_1 + \mbox{d}\theta^2_2 + \sin^2\theta_2 \mbox{d}\phi^2_2).
\label{t11}
\eea
One can compare this with the well-known Sasaki-Einstein metric \cite{Candelas:1989js}, 
\be
ds^2_{SE} = \tfrac{1}{9}\left(\mbox{d}\psi + \cos\theta_1 \mbox{d}\phi_1 + \cos\theta_2 \mbox{d}\phi_2\right)^2  
+\tfrac{1}{6}(\mbox{d}\theta^2_1 + \sin^2\theta_1 \mbox{d}\phi^2_1 + \mbox{d}\theta^2_2 + \sin^2\theta_2 \mbox{d}\phi^2_2).
\ee
The relative sizes of the Reeb vector, against the two-spheres within the K\"ahler-Einstein base,
are obviously different. Using the criteria given in Eq.~(2.8) of \cite{Candelas:1989js}, 
we see that the five-dimensional space in Eq.~(\ref{t11}) is  not Einstein. Topologically they
are both $S^2\times S^3$.

Imposing $\alpha=\beta=0$ is equivalent to setting $z_0=0$ 
in Eq.~(\ref{a1}).
Then we have $\{\sum_{i=1}^4 \,z_i^2 = 0\}$, i.e. a conifold singularity. On the field theory
side, due to the mapping $\Phi \rightarrow z_0$, we expect the dual operators would not contain any $\Phi$'s.

For explicit solutions we will again adopt the temporal gauge and set 
$
t=\kappa\tau.
$
Then we may write down the conserved charges as follows.
\bea
E &=& \frac{\sqrt{\lambda'}}{4}\kappa ,\\
R &=& \frac{3\sqrt{\lambda'}}{16}\int\frac{\mbox{d}^2\sigma}{(2\pi)^2}\biggl(\dot{\psi} + \cos\theta_1\dot{\phi_1} + \cos\theta_2\dot{\phi_2}\biggr)\,,\\
J_{A} &=& \frac{3\sqrt{\lambda'}}{128}\int\frac{\mbox{d}^2\sigma}{(2\pi)^2}\left[3\biggl(\dot{\psi} + \cos\theta_1\dot{\phi_1} + \cos\theta_2\dot{\phi_2}\biggr)\cos\theta_1 + 4\sin^2\theta_1\dot{\phi_1}\right]\,,\\
J_{B} &=& \frac{3\sqrt{\lambda'}}{128}\int\frac{\mbox{d}^2\sigma}{(2\pi)^2}\left[3\biggl(\dot{\psi} + \cos\theta_1\dot{\phi_1} + \cos\theta_2\dot{\phi_2}\biggr)\cos\theta_2 + 4\sin^2\theta_2\dot{\phi_2}\right]\, . 
\eea

We may find interesting solutions if 
\be
\phi_1 =0,\quad \theta_1=n\sigma_1,
\label{2pol}
\ee
with $n\in \mathbb{Z}$ 
and all other coordinates are functions of $(\tau,\sigma_2)$ only. 
Then 
remaining equations can be alternatively derived from an 
auxiliary Polyakov string action, moving within a squashed three-sphere
\be
\tfrac{1}{L^2}\mbox{d}s^2 = -\tfrac{1}{4}\mbox{d}t^2 + \tfrac{9}{64}\left(\mbox{d}\psi + \cos \theta \mbox{d}\phi\right)^2 + \tfrac{3}{16}(\mbox{d}\theta^2 + \sin^2\theta \mbox{d}\phi^2).\label{rmetric}
\ee
This metric is obtained as a subspace of Eq.~(\ref{t11}) if we suppress $\theta_1,\phi_1$ and 
rename $\sigma_2,\theta_2,\phi_2\rightarrow \sigma,\theta,\phi$ for brevity. The constraint equations
Eqs.~(\ref{constraint1}\ref{constraint2}) can be interpreted as Virasoro constraints if 
the string tension 
is given as
\be
T_P = \frac{1}{2\lambda^0} = \frac{\sqrt{3}}{4} nL T_2 .
\label{poly}
\ee

The study of classical rotating strings in the squashed three-sphere like Eq.~(\ref{rmetric})
has appeared several times
in the context of semiclassical string/membranes \cite{Kim:2003vn,Benvenuti:2008bd,Kim:2010ck}.
The equations are further reduced to a mechanical problem if we take the following ansatz.
\be
\psi=\omega\tau, \quad\phi=\nu\tau, \quad\theta=\theta(\sigma)\,. \label{ansatz}
\ee
The equations are either trivial or can be derived from the constraint condition 
\be
\theta'^2 + V(\theta) = {\cal E} , 
\ee
with
\be
{\cal E} = 
\tfrac{4\kappa^2}{3} , \quad
V(\theta) = \nu^2\sin^2\theta + \tfrac{3}{4}(\omega + \nu\cos\theta)^2\,.\label{potential}
\ee
Let us record here the conserved quantities for the ansatz Eq.~(\ref{2pol},\ref{ansatz}). We obtain 
\bea
\frac{E}{\sqrt{\lambda}} &=& \frac{e\kappa}{4} \,,\\
\frac{R}{\sqrt{\lambda}} &=& \frac{3e}{16}\int\frac{\mbox{d}\sigma}{2\pi}\bigl(\dot{\psi} + \cos\theta\dot{\phi}\bigr)\,,\\
\frac{J_A}{\sqrt{\lambda}} &=& 0 , \\
\frac{J_B}{\sqrt{\lambda}} &=& \frac{3e}{128}\int\frac{\mbox{d}\sigma}{2\pi}
\left[3\bigl(\dot{\psi} + \cos\theta\dot{\phi}\bigr)\cos\theta + 4\sin^2\theta\dot{\phi}\right]\, , 
\eea
where $e= \tfrac{\sqrt{3}n}{4}$.

The equation almost trivialises  if $\nu=0$. Then $V(\theta)$ is constant, periodicity 
demands $\theta_2=m\sigma_2 \,\,(m\in \mathbb{Z})$, and we have
{\it toroidal} rotating membranes with single spin. Let us just provide the dispersion relation here.
\be
E^2 = R^2 + \left(\tfrac{3nm\sqrt{\lambda}}{32}\right)^2, \quad J_A = J_B = 0 .
\ee
The dual operators are written in terms of $A_i,B_j$ only, with monopole operators when needed.
One can also expect that there must be the same number of $A_1$'s and $A_2$'s, because $J_A=0$. The same
argument applies to $B_i$'s.

In the following we consider the case of $\nu\neq 0$. The analysis 
will be very similar to that of \cite{Kim:2010ck,Kim:2003vn}. For cosmetic reasons, we define
$a=1,b=3/4,c=4/3,v=\kappa/\nu,\xi=\omega/\nu$. Then the following results can be applied to 
\be
V(\theta) = a\nu^2\sin^2\theta + b(\omega + \nu\cos\theta)^2, 
\ee
with ${\cal E}=c\kappa^2$.
\subsubsection{\label{3.2.1}Toroidal and cylindrical rotaing membranes with $\omega = 0$}
\begin{itemize}
\item Toroidal rotating membranes:

For simplicity, we first consider $\omega = 0$ case. The potential has a maximum value at $\theta = \pi/2$.
It $cv^2 < a$ the motion is vibrational, and the membrane is folded along $\theta$.
Due to the periodicity condition
$\theta(\sigma) = \theta(\sigma + 2\pi)$, we obtain
\be
\nu = \frac{2e}{\pi\sqrt{a-b}}\mathbf{K}(y)\,.
\ee
In the following we will use $y=\tfrac{cv^2 -b}{a-b}$. The nonvanishing charges are given as 
\bea
\frac{E}{\sqrt{\lambda}} &=& \,\frac{e}{\pi}\sqrt{\frac{y(a-b)+b}{c(a-b)}}\mathbf{K}(y)\,,\\
\frac{R}{\sqrt{\lambda}} &=& \frac{2e}{\sqrt{a-b}}\,,\\
\frac{J_B}{\sqrt{\lambda}} &=& \frac{e}{2\pi\sqrt{a-b}}\left(
4\mathbf{K}(y)-\mathbf{E}(y)\right)\,,
\eea
where $\mathbf{K}(y)$, $\mathbf{E}(y)$ are the first and second complete elliptic integrals. 
\item Cylindrical rotating membranes :

For $cv^2 \geq a$ or equivalently $y\geq 0$, we have rotating membranes cylindrical shape. Integration
of the equation of motion gives us the following results.
\bea
\nu &=& \frac{2}{\pi\sqrt{y(a-b)}}\mathbf{K}(1/y), \\
\frac{E}{\sqrt{\lambda}} &=& \frac{e}{\pi}\sqrt{\frac{y(a-b)+b}{cy(a-b)}}\mathbf{K}(1/y), \\
\frac{R}{\sqrt{\lambda}} &=& 0, \\
\frac{J_B}{\sqrt{\lambda}} &=& 
\frac{e}{2\pi\sqrt{y(a-b)}}
\left((3+y)\mathbf{K}(1/y)-y\mathbf{E}(1/y)\right) \, . 
\eea
\end{itemize}
In the limit $y\rightarrow 1$ the elliptic functions 
$\mathbf{K}(y)$ develope 
a logarithmic divergence, and the nonvanishing charges become large. One can 
express the dispersion 
relation as a series, using nome-$q$ expansion of elliptic integrals. 
The result is summarised
in the table~\ref{table1} .

\begin{table}
\centering
\begin{tabular}{c|c}
\hline\hline
  
    & $\text{Toroidal membrane}$
\\ [0.5ex]
\hline
${E}$ & $-\frac{3}{16\pi}$ln$q(1+2q+10q^2+\cdots)$
\\ [0.5ex]
$J_B$             & $-\frac{3\sqrt{3}}{64\pi}\{$ln$q(1+2q+8q^2+\cdots)+(\frac{1}{2}-2q+6q^2+\cdots)\}$
\\ [0.5ex]
$\text{Dispersion}$      & ${E}=\frac{4\sqrt{3}}{3}(J_{B}+\frac{3\sqrt{3}}{128\pi})-\frac{3}{8\pi}\mbox{exp}\,\biggl[-\frac{64\sqrt{3}\pi}{9}(J_B+\frac{3\sqrt{3}}{128\pi})\biggr]+\cdots$
\\ [1ex]
\hline
   & $\mbox{Cylindrical membrane}$
\\ [0.5ex]
\hline
${E}$ & $-\frac{3}{16\pi}$ln$q(1-2q+10q^2+\cdots)$
\\ [0.5ex]
$J_{B}$             & $-\frac{3\sqrt{3}}{64\pi}\{$ln$q(1-2q+8q^2+\cdots)+(\frac{1}{2}+2q+6q^2+\cdots)\}$
\\ [0.5ex]
$\mbox{Dispersion}$      & ${E}=\frac{4\sqrt{3}}{3}(J_{B}+\frac{3\sqrt{3}}{128\pi})+\frac{3}{8\pi}\mbox{exp}\,\biggl[-\frac{64\sqrt{3}\pi}{9}(J_B+\frac{3\sqrt{3}}{128\pi})\biggr]+\cdots$
\\ [1ex]
\hline
\end{tabular}
\caption{Q-series and dipersion relations for $\omega = 0$. (we put $\sqrt{\lambda} = 1$ and $n = 1$.)}
\label{table1}
\end{table}
\subsubsection{Multi-spin rotaing membranes with $\omega \neq 0$}
\label{3-2-2}
In this subsection we consider the general cylindrical motion of membrane with the potential~ (\ref{potential}).
From the constraint (\ref{constraint1}) and the periodicity condition $\theta(\sigma + 2\pi) = \theta(\sigma)$ we get
\bea
\nu = \frac{2}{\pi\sqrt{a-b}}\int^{\theta_0}_0 \frac{\mbox{d}\theta}{\sqrt{(\cos\theta - \alpha)(\cos\theta - \beta)}},
\eea
where $\theta_0 = \cos^{-1}\alpha$ and $\alpha$, $\beta$ ($\alpha > \beta$) are two roots of the quadratic equation
obtained from the constraint (\ref{constraint1}).
The energy and conserved chargs are expressed in terms of the elliptic integrals as follows.
\bea
\frac{E}{\sqrt{\lambda}} &=& \frac{e\,v}{\pi\sqrt{a-b}}\frac{1}{\sqrt{(1 + \alpha)(1 - \beta)}}\mathbf{K}(t),\\
\frac{R}{\sqrt{\lambda}} &=&
\frac{3e}{4\pi\sqrt{a-b}}\frac{1}{\sqrt{(1 + \alpha)(1 - \beta)}}\biggl[(\xi - 1)\mathbf{K}(t) + 2\mathbf{\Pi}(k,t)\biggr],\\
\frac{J_{B}}{\sqrt{\lambda}} &=&
\frac{3e}{16\pi\sqrt{a-b}}\frac{1}{\sqrt{(1 + \alpha)(1 - \beta)}}\biggl[(2-\frac{3}{2}\xi +\frac{\alpha}{2})\mathbf{K}(t)\nonumber\\
 &&+\left(3\xi - \frac{1}{2}(\alpha + \beta)\right)\mathbf{\Pi}(k,t)-\frac{1}{4}(1 +\alpha)(1-\beta)\mathbf{E}(t)\biggr],
\eea
where we have defined
\bea
k &=& -\frac{1 - \alpha}{1 + \alpha}, \\
t &=& \frac{(1 - \alpha)(1 + \beta)}{(1 + \alpha)(1 - \beta)}.
\eea
We would like to consider the case where the physical quantities become large. If we send
$v,-\xi$ $\rightarrow$ $\infty$, we will get regular
series expansions of $E$ and $J_{B}$ in terms of $J_R$. More concretely, we assume
\begin{equation}
c v^2 \sim b(\xi^2+2\varepsilon\xi+\cdots),
\end{equation}
with $-1 < \varepsilon < 1$. Then there exist folded spinning strings whose energy can be made arbitrarily
large. One easily sees $\alpha\rightarrow\varepsilon$, $\beta\rightarrow -\infty$ and $k,t\rightarrow
\frac{\varepsilon-1}{\varepsilon+1}$ in this limit.
We will obtain series expansions of the following form,
\begin{eqnarray}
E            &=& a_1R+a_2\frac{\lambda}{R}+a_3\frac{\lambda^2}{R^3}+\cdots,  \\
J_{B}    &=& b_1R+b_2\frac{\lambda}{R}+b_3\frac{\lambda^2}{R^3}+\cdots.
\end{eqnarray}
We have computed first few coefficients,
\begin{eqnarray}
a_1   &=&  1,  \label{ub} \\
a_2   &=&  \frac{9}{128\pi^2}\,\mathbf{K}(-z)\,(\mathbf{E}(-z)-\mathbf{K}(-z)), 
\label{el1}\\
b_1   &=&  \frac{3}{8}-\frac{3}{4(z+1)}\frac{\mathbf{E}(-z)}{\mathbf{K}(-z)},
\label{el2}
\end{eqnarray}
where
\begin{equation}
z=\frac{1-\varepsilon}{1+\varepsilon}.
\end{equation}

Let us consider the dual operators. Eq.~(\ref{ub}) implies that
the unitarity bound is asymptotically saturated, so we expect the
dual operators are made of $A_i,B_i$ and not their complex conjugates.
Since $J_A=0$, it is clear that there are the same number of {\it spin-ups}
$A_1$ and {\it spin-downs} $A_2$. For the composition of $(B_1,B_2)$, we have
a complicated implicit relation given by the series. 

The appearance of 
elliptic integrals is in fact not unfamiliar in the spectrum of 
integrable spin chain. For the duality of $\cN=4$ super Yang-Mills theory
and IIB strings in $\mbox{AdS}_5\times S^5$, one can push the computations
on both sides of the duality. The coefficient $a_2$ encodes 
${\cal O}(\lambda)$, i.e. one-loop computation which is conveniently
summarised as spin system with nearest-neighbor interactions. The spinning
string result and the Bethe ansatz result show exact match \cite{Beisert:2003ea}.
Although we do not have a viable field theory result since the ${\cal A}_1$
quiver theory 
 is strongly
coupled, it would be very nice if Eqs.~(\ref{el1}\ref{el2}) can be 
derived from an integrable $SO(5)$ spin chain.

\subsection{Giant magnon and single spike solutions}
On the string theory side, giant magnons and spiky string solutions are
rotating open strings with infinite energy and angular momenta. For the giant
magnon solutions on $S^2$ the end points are on the equator and carry
infinite momenta.

Benvenuti and Tonni studied giant magnons and single spikes in squashed
three-sphere \cite{Benvenuti:2008bd}. Since we have reduced the rotating
membrane problem into a Polyakov string moving in the squashed three-sphere, 
the results of \cite{Benvenuti:2008bd} are directly applicable here. Instead of
repeating all the derivations, we will very briefly explain what kind of
solutions we are after, and present the result.

For giant magnons and single spikes, one introduces the following ansatz for
string motion in Eq.~(\ref{rmetric}). 
\bea
t = \kappa\tau, \quad \theta = \theta(x),\quad \psi = \omega\tau + \Psi(x), \quad
\phi = \nu\tau + \Phi(x),\label{mansatz}
\eea
where $x=\beta\sigma - \gamma\tau$ and $\beta,\gamma$ are constants. 

The equation for $t$ is still trivial. One then looks at the equations for
 $\psi,\phi$ and find they are nontrivial, unlike the spinning strings we have 
considered so far. Fortunately they can be  integrated once, and one can express
$\Psi',\Phi'$ as functions of $\theta$. If we plug those relations into 
the equation of motion for $\theta$, we obtain a mechanical system with
$\theta'^2 + V(\theta)=0$. Giant magnons and spiky strings are characterised by 
the condition
$V(\theta)\propto -(\theta-\pi)^2$ which implies it takes infinitely long to 
arrive at the {\it turning point} $\theta=\pi$. In terms of strings, this 
means the open string end points lie at $\theta=\pi$. The giant magnons
are U-shaped and characterized by $\Delta \phi$  which is the difference of $\phi$ 
coordinate at the two end points. With spiky strings $\Delta\phi$ is divergent
as well as energy.

The analysis of \cite{Benvenuti:2008bd} is for general squashed three-sphere
with $SU(2)\times U(1)$ symmetry. We can translate their result for our setting
with simple re-scalings. We will just record the dispersion relations. For 
simplicity we set $\sqrt\lambda=1$. 

For giant magnon type solutions, $E,R,J_B$ are all infinite but 
the linear combinations $E-R,E-8J_B/3$ are finite. The dispersion relations is 
\bea
E - \frac{8}{3}{J_{B}} = \frac{n}{2}\frac{\cos(4(E-R)/3n) -
\cos\Delta\varphi}{\sin(4(E-R)/3n)} . 
\eea

 For a single spike, $\Delta\phi,E$ are divergent but $R,J_B$ are finite.
The dispersion relation is 
\bea
J_{B} = \frac{3n}{16} \frac{\cos(8E/3n - \Delta\varphi) - \cos(4R/3n)}{\sin(4R/3n)}.
\eea
\section{\label{4}DISCUSSION}
We have studied rotating membranes in nono-toric
Sasaki-Einstein manifold $V_{5,2}$. The metric is still homogeneous, but
more complicated than $Q^{1,1,1}$ or $M^{1,1,1}$ which are toric. 
At first sight, probably only the $SU(2)\times SU(2)\times U(1)_R$ part of the 
isometry is manifest.

The main task
in this article is to construct explicit solutions and provide their field
theory interpretations as long operators. We have five {\it alphabets}, i.e.
chiral multiplets to be used in the construction of chiral primaries. They
are $\Phi,A_1,A_2,B_1,B_2$ and the last four fields parametrize the manifest $SU(2)\times SU(2)$ symmetry. We managed to find 
a couple of wrapped membrane solutions which are dual to operators containing
$\Phi$'s. More sophisticated solutions are constructed in the $T^{1,1}$ subsector.
We expect the dual operators are then made of $A_i,B_i$ and do not contain $\Phi$'s.

Eventually we would like to have some field theory side computations and compare
them against our results in this paper. Since the dual field theory is at 
a nontrivial fixed point of the renormalization group, conventional field theory
techniques are not applicable. According to the standard AdS/CFT dictionary, 
the classical membrane results correspond to all-order results in the planar
limit. It would be nice if we could derive from the membrane solutions or the Hamiltonian operator in the 
radial quantization prescription of conformal field theory. 

\acknowledgments
N. Kim is very grateful to D. Martelli for helpful discussions and comments.
J. Kim has been partly supported by Institute for the Early Universe (IEU) at Ewha university 
during the later stages of this work. 
This research is supported by the National Research Foundation of Korea (NRF) funded by 
the Korea government (MEST) with the 
grant No. 2009-0085995
and also partly through the Center for Quantum Spactime (CQUeST) of Sogang University 
with grant No. R11-2005-021.

\appendix

\section{Useful formulas for elliptic integrals}
 In this appendix we present some definitions and relations which is needed to obtain our results in this note. The complete elliptic integrals of the first and the second is defined as below:
\begin{align}
\mathbf{K}(m)&= \int_0^{\pi/2} \frac{d\phi}{\sqrt{1-m\,\text{sin}^2 \phi}},\\
\mathbf{E}(m)&= \int_0^{\pi/2} \sqrt{1-m\,\text{sin}^2\phi}\,d\phi ,\\
\mathbf{\Pi}(k,m)&=\int_0^{\pi/2} \frac{d\phi}{(1-k\,\text{sin}^2\phi) \sqrt{1-m\,\text{sin}^2\phi}}.
\end{align}
 We often use the following relation
\begin{equation}
I(n)=\int_\alpha^1 \frac{s^n\,ds}{\sqrt{(s-\alpha)(s-\beta)(1-s^2)}}
\,\,\,\,\,\,\,\,\,\,\,\,(\beta<\alpha<1),\end{equation}
which is used to derive the following relations
\begin{align}
I(0)&=\frac{2}{\sqrt{(1+\alpha)(1-\beta)}}\mathbf{K}(t),\\
I(1)&=\frac{2}{\sqrt{(1+\alpha)(1-\beta)}}\biggl[ 2\mathbf{\Pi}(k,t)-\mathbf{K}(t) \biggr],\\
I(2)&=\frac{1}{\sqrt{(1+\alpha)(1-\beta)}}\biggl[2(\alpha +\beta)\mathbf{\Pi}(k,t)-2\alpha\mathbf{K}(t)+(1+\alpha)(1-\beta)\mathbf{E}(t) \biggr],
\end{align}
where
\begin{equation}
k=-\frac{1-\alpha}{1+\alpha},\,\,\,\,\,\, t=\frac{(1-\alpha)(1+\beta)}{(1+\alpha)(1-\beta)}.
\end{equation}
 In order to study the elliptic integral near the logarithmic singularities, it is convenient to use the q-series and defined as
\begin{align}
q &\equiv\text{exp}[-\pi\mathbf{K}(1-m)/\mathbf{K}(m)] \\
  &=\frac{m}{16} + 8\biggl(\frac{m}{16}\biggr)^2 + 84\biggl(\frac{m}{16}\biggr)^3 +\cdots
\end{align}
 Inverting the above relation, one obtains
\begin{equation}
m=16(q-8q^2+44q^3-192q^4+\cdots)
\end{equation}
Then one has the following alternative expansions of elliptic integrals.
\begin{align}
\mathbf{K}(m)&=\frac{\pi}{2}(1+4q+4q^2+4q^4+\cdots),\\
\mathbf{E}(m)&=\frac{\pi}{2}(1-4q+20q^2-64q^3+\cdots).
\end{align}
And more importantly,
\begin{align}
\mathbf{K}(1-m)&=-\frac{\text{ln}q}{2}(1+4q+4q^2+4q^4+\cdots),\\
\mathbf{E}(1-m)&=(1-4q+12q^2-32q^3+\cdots)-4q\,\text{ln}q(1-2q+8q^2+\cdots).
\end{align}
To get the expansion of $\mathbf{E}(1-m)$ from the above, it is convenient to use the Legendre's relation,
\begin{equation}
\mathbf{E}(m)\mathbf{K}(1-m)+\mathbf{E}(1-m)\mathbf{K}(m)-\mathbf{K}(m)\mathbf{K}(1-m)=\frac{\pi}{2}
\end{equation}

\bibliography{v52}{}

\begin{thebibliography}{32}%
\makeatletter
\providecommand \@ifxundefined [1]{%
 \@ifx{#1\undefined}
}%
\providecommand \@ifnum [1]{%
 \ifnum #1\expandafter \@firstoftwo
 \else \expandafter \@secondoftwo
 \fi
}%
\providecommand \@ifx [1]{%
 \ifx #1\expandafter \@firstoftwo
 \else \expandafter \@secondoftwo
 \fi
}%
\providecommand \natexlab [1]{#1}%
\providecommand \enquote  [1]{``#1''}%
\providecommand \bibnamefont  [1]{#1}%
\providecommand \bibfnamefont [1]{#1}%
\providecommand \citenamefont [1]{#1}%
\providecommand \href@noop [0]{\@secondoftwo}%
\providecommand \href [0]{\begingroup \@sanitize@url \@href}%
\providecommand \@href[1]{\@@startlink{#1}\@@href}%
\providecommand \@@href[1]{\endgroup#1\@@endlink}%
\providecommand \@sanitize@url [0]{\catcode `\\12\catcode `\$12\catcode
  `\&12\catcode `\#12\catcode `\^12\catcode `\_12\catcode `\%12\relax}%
\providecommand \@@startlink[1]{}%
\providecommand \@@endlink[0]{}%
\providecommand \url  [0]{\begingroup\@sanitize@url \@url }%
\providecommand \@url [1]{\endgroup\@href {#1}{\urlprefix }}%
\providecommand \urlprefix  [0]{URL }%
\providecommand \Eprint [0]{\href }%
\@ifxundefined \urlstyle {%
  \providecommand \doi  [0]{\begingroup \@sanitize@url \@doi}%
  \providecommand \@doi [1]{\endgroup \@@startlink {\doibase
  #1}doi:\discretionary {}{}{}#1\@@endlink }%
}{%
  \providecommand \doi  [0]{doi:\discretionary{}{}{}\begingroup
  \urlstyle{rm}\Url }%
}%
\providecommand \doibase [0]{http://dx.doi.org/}%
\providecommand \Doi [0]{\begingroup \@sanitize@url \@Doi }%
\providecommand \@Doi  [1]{\endgroup\@@startlink{\doibase#1}\@@Doi}%
\providecommand \@@Doi [1]{#1\@@endlink}%
\providecommand \selectlanguage [0]{\@gobble}%
\providecommand \bibinfo  [0]{\@secondoftwo}%
\providecommand \bibfield  [0]{\@secondoftwo}%
\providecommand \translation [1]{[#1]}%
\providecommand \BibitemOpen [0]{}%
\providecommand \bibitemStop [0]{}%
\providecommand \bibitemNoStop [0]{.\EOS\space}%
\providecommand \EOS [0]{\spacefactor3000\relax}%
\providecommand \BibitemShut  [1]{\csname bibitem#1\endcsname}%
\bibitem [{\citenamefont {Aharony}\ \emph {et~al.}(2008)\citenamefont
  {Aharony}, \citenamefont {Bergman}, \citenamefont {Jafferis},\ and\
  \citenamefont {Maldacena}}]{Aharony:2008ug}%
  \BibitemOpen
  \bibfield  {author} {\bibinfo {author} {\bibfnamefont {O.}~\bibnamefont
  {Aharony}}, \bibinfo {author} {\bibfnamefont {O.}~\bibnamefont {Bergman}},
  \bibinfo {author} {\bibfnamefont {D.~L.}\ \bibnamefont {Jafferis}}, \ and\
  \bibinfo {author} {\bibfnamefont {J.}~\bibnamefont {Maldacena}},\ }\Doi
  {10.1088/1126-6708/2008/10/091} {\bibfield  {journal} {\bibinfo  {journal}
  {JHEP},\ }\textbf {\bibinfo {volume} {10}},\ \bibinfo {pages} {091} (\bibinfo
  {year} {2008})},\ \Eprint {http://arxiv.org/abs/0806.1218} {arXiv:0806.1218
  [hep-th]} \BibitemShut {NoStop}%
\bibitem [{\citenamefont {Imamura}\ and\ \citenamefont
  {Kimura}(2008){\natexlab{a}}}]{Imamura:2008nn}%
  \BibitemOpen
  \bibfield  {author} {\bibinfo {author} {\bibfnamefont {Y.}~\bibnamefont
  {Imamura}}\ and\ \bibinfo {author} {\bibfnamefont {K.}~\bibnamefont
  {Kimura}},\ }\Doi {10.1143/PTP.120.509} {\bibfield  {journal} {\bibinfo
  {journal} {Prog. Theor. Phys.},\ }\textbf {\bibinfo {volume} {120}},\
  \bibinfo {pages} {509} (\bibinfo {year} {2008}{\natexlab{a}})},\ \Eprint
  {http://arxiv.org/abs/0806.3727} {arXiv:0806.3727 [hep-th]} \BibitemShut
  {NoStop}%
\bibitem [{\citenamefont {Martelli}\ and\ \citenamefont
  {Sparks}(2008)}]{Martelli:2008si}%
  \BibitemOpen
  \bibfield  {author} {\bibinfo {author} {\bibfnamefont {D.}~\bibnamefont
  {Martelli}}\ and\ \bibinfo {author} {\bibfnamefont {J.}~\bibnamefont
  {Sparks}},\ }\Doi {10.1103/PhysRevD.78.126005} {\bibfield  {journal}
  {\bibinfo  {journal} {Phys. Rev.},\ }\textbf {\bibinfo {volume} {D78}},\
  \bibinfo {pages} {126005} (\bibinfo {year} {2008})},\ \Eprint
  {http://arxiv.org/abs/0808.0912} {arXiv:0808.0912 [hep-th]} \BibitemShut
  {NoStop}%
\bibitem [{\citenamefont {Hanany}\ and\ \citenamefont
  {Zaffaroni}(2008)}]{Hanany:2008cd}%
  \BibitemOpen
  \bibfield  {author} {\bibinfo {author} {\bibfnamefont {A.}~\bibnamefont
  {Hanany}}\ and\ \bibinfo {author} {\bibfnamefont {A.}~\bibnamefont
  {Zaffaroni}},\ }\Doi {10.1088/1126-6708/2008/10/111} {\bibfield  {journal}
  {\bibinfo  {journal} {JHEP},\ }\textbf {\bibinfo {volume} {10}},\ \bibinfo
  {pages} {111} (\bibinfo {year} {2008})},\ \Eprint
  {http://arxiv.org/abs/0808.1244} {arXiv:0808.1244 [hep-th]} \BibitemShut
  {NoStop}%
\bibitem [{\citenamefont {Imamura}\ and\ \citenamefont
  {Kimura}(2008){\natexlab{b}}}]{Imamura:2008qs}%
  \BibitemOpen
  \bibfield  {author} {\bibinfo {author} {\bibfnamefont {Y.}~\bibnamefont
  {Imamura}}\ and\ \bibinfo {author} {\bibfnamefont {K.}~\bibnamefont
  {Kimura}},\ }\Doi {10.1088/1126-6708/2008/10/114} {\bibfield  {journal}
  {\bibinfo  {journal} {JHEP},\ }\textbf {\bibinfo {volume} {10}},\ \bibinfo
  {pages} {114} (\bibinfo {year} {2008}{\natexlab{b}})},\ \Eprint
  {http://arxiv.org/abs/0808.4155} {arXiv:0808.4155 [hep-th]} \BibitemShut
  {NoStop}%
\bibitem [{\citenamefont {Hanany}\ \emph {et~al.}(2009)\citenamefont {Hanany},
  \citenamefont {Vegh},\ and\ \citenamefont {Zaffaroni}}]{Hanany:2008fj}%
  \BibitemOpen
  \bibfield  {author} {\bibinfo {author} {\bibfnamefont {A.}~\bibnamefont
  {Hanany}}, \bibinfo {author} {\bibfnamefont {D.}~\bibnamefont {Vegh}}, \ and\
  \bibinfo {author} {\bibfnamefont {A.}~\bibnamefont {Zaffaroni}},\ }\Doi
  {10.1088/1126-6708/2009/03/012} {\bibfield  {journal} {\bibinfo  {journal}
  {JHEP},\ }\textbf {\bibinfo {volume} {03}},\ \bibinfo {pages} {012} (\bibinfo
  {year} {2009})},\ \Eprint {http://arxiv.org/abs/0809.1440} {arXiv:0809.1440
  [hep-th]} \BibitemShut {NoStop}%
\bibitem [{\citenamefont {Franco}\ \emph {et~al.}(2008)\citenamefont {Franco},
  \citenamefont {Hanany}, \citenamefont {Park},\ and\ \citenamefont
  {Rodriguez-Gomez}}]{Franco:2008um}%
  \BibitemOpen
  \bibfield  {author} {\bibinfo {author} {\bibfnamefont {S.}~\bibnamefont
  {Franco}}, \bibinfo {author} {\bibfnamefont {A.}~\bibnamefont {Hanany}},
  \bibinfo {author} {\bibfnamefont {J.}~\bibnamefont {Park}}, \ and\ \bibinfo
  {author} {\bibfnamefont {D.}~\bibnamefont {Rodriguez-Gomez}},\ }\Doi
  {10.1088/1126-6708/2008/12/110} {\bibfield  {journal} {\bibinfo  {journal}
  {JHEP},\ }\textbf {\bibinfo {volume} {12}},\ \bibinfo {pages} {110} (\bibinfo
  {year} {2008})},\ \Eprint {http://arxiv.org/abs/0809.3237} {arXiv:0809.3237
  [hep-th]} \BibitemShut {NoStop}%
\bibitem [{\citenamefont {Franco}\ \emph {et~al.}(2009)\citenamefont {Franco},
  \citenamefont {Klebanov},\ and\ \citenamefont
  {Rodriguez-Gomez}}]{Franco:2009sp}%
  \BibitemOpen
  \bibfield  {author} {\bibinfo {author} {\bibfnamefont {S.}~\bibnamefont
  {Franco}}, \bibinfo {author} {\bibfnamefont {I.~R.}\ \bibnamefont
  {Klebanov}}, \ and\ \bibinfo {author} {\bibfnamefont {D.}~\bibnamefont
  {Rodriguez-Gomez}},\ }\Doi {10.1088/1126-6708/2009/08/033} {\bibfield
  {journal} {\bibinfo  {journal} {JHEP},\ }\textbf {\bibinfo {volume} {08}},\
  \bibinfo {pages} {033} (\bibinfo {year} {2009})},\ \Eprint
  {http://arxiv.org/abs/0903.3231} {arXiv:0903.3231 [hep-th]} \BibitemShut
  {NoStop}%
\bibitem [{\citenamefont {Aganagic}(2009)}]{Aganagic:2009zk}%
  \BibitemOpen
  \bibfield  {author} {\bibinfo {author} {\bibfnamefont {M.}~\bibnamefont
  {Aganagic}},\ }\href@noop {} { (\bibinfo {year} {2009})},\ \Eprint
  {http://arxiv.org/abs/0905.3415} {arXiv:0905.3415 [hep-th]} \BibitemShut
  {NoStop}%
\bibitem [{\citenamefont {Martelli}\ and\ \citenamefont
  {Sparks}(2009)}]{Martelli:2009ga}%
  \BibitemOpen
  \bibfield  {author} {\bibinfo {author} {\bibfnamefont {D.}~\bibnamefont
  {Martelli}}\ and\ \bibinfo {author} {\bibfnamefont {J.}~\bibnamefont
  {Sparks}},\ }\Doi {10.1088/1126-6708/2009/12/017} {\bibfield  {journal}
  {\bibinfo  {journal} {JHEP},\ }\textbf {\bibinfo {volume} {12}},\ \bibinfo
  {pages} {017} (\bibinfo {year} {2009})},\ \Eprint
  {http://arxiv.org/abs/0909.2036} {arXiv:0909.2036 [hep-th]} \BibitemShut
  {NoStop}%
\bibitem [{\citenamefont {Elitzur}\ \emph {et~al.}(1997)\citenamefont
  {Elitzur}, \citenamefont {Giveon},\ and\ \citenamefont
  {Kutasov}}]{Elitzur:1997fh}%
  \BibitemOpen
  \bibfield  {author} {\bibinfo {author} {\bibfnamefont {S.}~\bibnamefont
  {Elitzur}}, \bibinfo {author} {\bibfnamefont {A.}~\bibnamefont {Giveon}}, \
  and\ \bibinfo {author} {\bibfnamefont {D.}~\bibnamefont {Kutasov}},\ }\Doi
  {10.1016/S0370-2693(97)00375-4} {\bibfield  {journal} {\bibinfo  {journal}
  {Phys. Lett.},\ }\textbf {\bibinfo {volume} {B400}},\ \bibinfo {pages} {269}
  (\bibinfo {year} {1997})},\ \Eprint {http://arxiv.org/abs/hep-th/9702014}
  {arXiv:hep-th/9702014} \BibitemShut {NoStop}%
\bibitem [{\citenamefont {Franco}\ \emph {et~al.}(2006)\citenamefont {Franco}
  \emph {et~al.}}]{Franco:2005sm}%
  \BibitemOpen
  \bibfield  {author} {\bibinfo {author} {\bibfnamefont {S.}~\bibnamefont
  {Franco}} \emph {et~al.},\ }\href@noop {} {\bibfield  {journal} {\bibinfo
  {journal} {JHEP},\ }\textbf {\bibinfo {volume} {01}},\ \bibinfo {pages} {128}
  (\bibinfo {year} {2006})},\ \Eprint {http://arxiv.org/abs/hep-th/0505211}
  {arXiv:hep-th/0505211} \BibitemShut {NoStop}%
\bibitem [{\citenamefont {Davey}\ \emph {et~al.}(2009)\citenamefont {Davey},
  \citenamefont {Hanany}, \citenamefont {Mekareeya},\ and\ \citenamefont
  {Torri}}]{Davey:2009et}%
  \BibitemOpen
  \bibfield  {author} {\bibinfo {author} {\bibfnamefont {J.}~\bibnamefont
  {Davey}}, \bibinfo {author} {\bibfnamefont {A.}~\bibnamefont {Hanany}},
  \bibinfo {author} {\bibfnamefont {N.}~\bibnamefont {Mekareeya}}, \ and\
  \bibinfo {author} {\bibfnamefont {G.}~\bibnamefont {Torri}},\ }\href@noop {}
  { (\bibinfo {year} {2009})},\ \Eprint {http://arxiv.org/abs/0910.4962}
  {arXiv:0910.4962 [hep-th]} \BibitemShut {NoStop}%
\bibitem [{\citenamefont {Gubser}\ \emph {et~al.}(2002)\citenamefont {Gubser},
  \citenamefont {Klebanov},\ and\ \citenamefont {Polyakov}}]{Gubser:2002tv}%
  \BibitemOpen
  \bibfield  {author} {\bibinfo {author} {\bibfnamefont {S.~S.}\ \bibnamefont
  {Gubser}}, \bibinfo {author} {\bibfnamefont {I.~R.}\ \bibnamefont
  {Klebanov}}, \ and\ \bibinfo {author} {\bibfnamefont {A.~M.}\ \bibnamefont
  {Polyakov}},\ }\Doi {10.1016/S0550-3213(02)00373-5} {\bibfield  {journal}
  {\bibinfo  {journal} {Nucl. Phys.},\ }\textbf {\bibinfo {volume} {B636}},\
  \bibinfo {pages} {99} (\bibinfo {year} {2002})},\ \Eprint
  {http://arxiv.org/abs/hep-th/0204051} {arXiv:hep-th/0204051} \BibitemShut
  {NoStop}%
\bibitem [{\citenamefont {Bozhilov}(2005)}]{Bozhilov:2005ew}%
  \BibitemOpen
  \bibfield  {author} {\bibinfo {author} {\bibfnamefont {P.}~\bibnamefont
  {Bozhilov}},\ }\href@noop {} {\bibfield  {journal} {\bibinfo  {journal}
  {JHEP},\ }\textbf {\bibinfo {volume} {08}},\ \bibinfo {pages} {087} (\bibinfo
  {year} {2005})},\ \Eprint {http://arxiv.org/abs/hep-th/0507149}
  {arXiv:hep-th/0507149} \BibitemShut {NoStop}%
\bibitem [{\citenamefont {Bozhilov}(2006){\natexlab{a}}}]{Bozhilov:2005xs}%
  \BibitemOpen
  \bibfield  {author} {\bibinfo {author} {\bibfnamefont {P.}~\bibnamefont
  {Bozhilov}},\ }\href@noop {} {\bibfield  {journal} {\bibinfo  {journal}
  {JHEP},\ }\textbf {\bibinfo {volume} {03}},\ \bibinfo {pages} {001} (\bibinfo
  {year} {2006}{\natexlab{a}})},\ \Eprint {http://arxiv.org/abs/hep-th/0511253}
  {arXiv:hep-th/0511253} \BibitemShut {NoStop}%
\bibitem [{\citenamefont {Bozhilov}(2006){\natexlab{b}}}]{Bozhilov:2006gh}%
  \BibitemOpen
  \bibfield  {author} {\bibinfo {author} {\bibfnamefont {P.}~\bibnamefont
  {Bozhilov}},\ }\href@noop {} { (\bibinfo {year} {2006}{\natexlab{b}})},\
  \Eprint {http://arxiv.org/abs/hep-th/0612175} {arXiv:hep-th/0612175}
  \BibitemShut {NoStop}%
\bibitem [{\citenamefont {Bozhilov}(2007){\natexlab{a}}}]{Bozhilov:2007km}%
  \BibitemOpen
  \bibfield  {author} {\bibinfo {author} {\bibfnamefont {P.}~\bibnamefont
  {Bozhilov}},\ }\Doi {10.1103/PhysRevD.76.106003} {\bibfield  {journal}
  {\bibinfo  {journal} {Phys. Rev.},\ }\textbf {\bibinfo {volume} {D76}},\
  \bibinfo {pages} {106003} (\bibinfo {year} {2007}{\natexlab{a}})},\ \Eprint
  {http://arxiv.org/abs/0706.1443} {arXiv:0706.1443 [hep-th]} \BibitemShut
  {NoStop}%
\bibitem [{\citenamefont {Bozhilov}(2007){\natexlab{b}}}]{Bozhilov:2007mb}%
  \BibitemOpen
  \bibfield  {author} {\bibinfo {author} {\bibfnamefont {P.}~\bibnamefont
  {Bozhilov}},\ }\Doi {10.1088/1126-6708/2007/08/073} {\bibfield  {journal}
  {\bibinfo  {journal} {JHEP},\ }\textbf {\bibinfo {volume} {08}},\ \bibinfo
  {pages} {073} (\bibinfo {year} {2007}{\natexlab{b}})},\ \Eprint
  {http://arxiv.org/abs/0704.3082} {arXiv:0704.3082 [hep-th]} \BibitemShut
  {NoStop}%
\bibitem [{\citenamefont {Bozhilov}(2008)}]{Bozhilov:2007bi}%
  \BibitemOpen
  \bibfield  {author} {\bibinfo {author} {\bibfnamefont {P.}~\bibnamefont
  {Bozhilov}},\ }\Doi {10.1002/prop.200710508} {\bibfield  {journal} {\bibinfo
  {journal} {Fortsch. Phys.},\ }\textbf {\bibinfo {volume} {56}},\ \bibinfo
  {pages} {373} (\bibinfo {year} {2008})},\ \Eprint
  {http://arxiv.org/abs/0711.1524} {arXiv:0711.1524 [hep-th]} \BibitemShut
  {NoStop}%
\bibitem [{\citenamefont {Bozhilov}\ and\ \citenamefont
  {Rashkov}(2008)}]{Bozhilov:2007wn}%
  \BibitemOpen
  \bibfield  {author} {\bibinfo {author} {\bibfnamefont {P.}~\bibnamefont
  {Bozhilov}}\ and\ \bibinfo {author} {\bibfnamefont {R.~C.}\ \bibnamefont
  {Rashkov}},\ }\Doi {10.1016/j.nuclphysb.2007.10.004} {\bibfield  {journal}
  {\bibinfo  {journal} {Nucl. Phys.},\ }\textbf {\bibinfo {volume} {B794}},\
  \bibinfo {pages} {429} (\bibinfo {year} {2008})},\ \Eprint
  {http://arxiv.org/abs/0708.0325} {arXiv:0708.0325 [hep-th]} \BibitemShut
  {NoStop}%
\bibitem [{\citenamefont {Wen}(2008)}]{Wen:2007az}%
  \BibitemOpen
  \bibfield  {author} {\bibinfo {author} {\bibfnamefont {W.-Y.}\ \bibnamefont
  {Wen}},\ }\Doi {10.1016/j.nuclphysb.2007.09.027} {\bibfield  {journal}
  {\bibinfo  {journal} {Nucl. Phys.},\ }\textbf {\bibinfo {volume} {B791}},\
  \bibinfo {pages} {164} (\bibinfo {year} {2008})},\ \Eprint
  {http://arxiv.org/abs/0705.2634} {arXiv:0705.2634 [hep-th]} \BibitemShut
  {NoStop}%
\bibitem [{\citenamefont {Ahn}\ and\ \citenamefont
  {Bozhilov}(2008)}]{Ahn:2008gd}%
  \BibitemOpen
  \bibfield  {author} {\bibinfo {author} {\bibfnamefont {C.}~\bibnamefont
  {Ahn}}\ and\ \bibinfo {author} {\bibfnamefont {P.}~\bibnamefont {Bozhilov}},\
  }\Doi {10.1088/1126-6708/2008/08/054} {\bibfield  {journal} {\bibinfo
  {journal} {JHEP},\ }\textbf {\bibinfo {volume} {08}},\ \bibinfo {pages} {054}
  (\bibinfo {year} {2008})},\ \Eprint {http://arxiv.org/abs/0807.0566}
  {arXiv:0807.0566 [hep-th]} \BibitemShut {NoStop}%
\bibitem [{\citenamefont {Kim}\ \emph {et~al.}(2010)\citenamefont {Kim},
  \citenamefont {Kim},\ and\ \citenamefont {Lee}}]{Kim:2010ck}%
  \BibitemOpen
  \bibfield  {author} {\bibinfo {author} {\bibfnamefont {J.}~\bibnamefont
  {Kim}}, \bibinfo {author} {\bibfnamefont {N.}~\bibnamefont {Kim}}, \ and\
  \bibinfo {author} {\bibfnamefont {J.~H.}\ \bibnamefont {Lee}},\ }\href@noop
  {} { (\bibinfo {year} {2010})},\ \Eprint {http://arxiv.org/abs/1001.2902}
  {arXiv:1001.2902 [hep-th]} \BibitemShut {NoStop}%
\bibitem [{\citenamefont {Ceresole}\ \emph {et~al.}(2000)\citenamefont
  {Ceresole}, \citenamefont {Dall'Agata}, \citenamefont {D'Auria},\ and\
  \citenamefont {Ferrara}}]{Ceresole:1999zg}%
  \BibitemOpen
  \bibfield  {author} {\bibinfo {author} {\bibfnamefont {A.}~\bibnamefont
  {Ceresole}}, \bibinfo {author} {\bibfnamefont {G.}~\bibnamefont
  {Dall'Agata}}, \bibinfo {author} {\bibfnamefont {R.}~\bibnamefont {D'Auria}},
  \ and\ \bibinfo {author} {\bibfnamefont {S.}~\bibnamefont {Ferrara}},\
  }\href@noop {} {\bibfield  {journal} {\bibinfo  {journal} {JHEP},\ }\textbf
  {\bibinfo {volume} {03}},\ \bibinfo {pages} {011} (\bibinfo {year} {2000})},\
  \Eprint {http://arxiv.org/abs/hep-th/9912107} {arXiv:hep-th/9912107}
  \BibitemShut {NoStop}%
\bibitem [{\citenamefont {Hofman}\ and\ \citenamefont
  {Maldacena}(2006)}]{Hofman:2006xt}%
  \BibitemOpen
  \bibfield  {author} {\bibinfo {author} {\bibfnamefont {D.~M.}\ \bibnamefont
  {Hofman}}\ and\ \bibinfo {author} {\bibfnamefont {J.~M.}\ \bibnamefont
  {Maldacena}},\ }\Doi {10.1088/0305-4470/39/41/S17} {\bibfield  {journal}
  {\bibinfo  {journal} {J. Phys.},\ }\textbf {\bibinfo {volume} {A39}},\
  \bibinfo {pages} {13095} (\bibinfo {year} {2006})},\ \Eprint
  {http://arxiv.org/abs/hep-th/0604135} {arXiv:hep-th/0604135} \BibitemShut
  {NoStop}%
\bibitem [{\citenamefont {Kruczenski}(2005)}]{Kruczenski:2004wg}%
  \BibitemOpen
  \bibfield  {author} {\bibinfo {author} {\bibfnamefont {M.}~\bibnamefont
  {Kruczenski}},\ }\href@noop {} {\bibfield  {journal} {\bibinfo  {journal}
  {JHEP},\ }\textbf {\bibinfo {volume} {08}},\ \bibinfo {pages} {014} (\bibinfo
  {year} {2005})},\ \Eprint {http://arxiv.org/abs/hep-th/0410226}
  {arXiv:hep-th/0410226} \BibitemShut {NoStop}%
\bibitem [{\citenamefont {Bergman}\ and\ \citenamefont
  {Herzog}(2002)}]{Bergman:2001qi}%
  \BibitemOpen
  \bibfield  {author} {\bibinfo {author} {\bibfnamefont {A.}~\bibnamefont
  {Bergman}}\ and\ \bibinfo {author} {\bibfnamefont {C.~P.}\ \bibnamefont
  {Herzog}},\ }\href@noop {} {\bibfield  {journal} {\bibinfo  {journal}
  {JHEP},\ }\textbf {\bibinfo {volume} {01}},\ \bibinfo {pages} {030} (\bibinfo
  {year} {2002})},\ \Eprint {http://arxiv.org/abs/hep-th/0108020}
  {arXiv:hep-th/0108020} \BibitemShut {NoStop}%
\bibitem [{\citenamefont {Candelas}\ and\ \citenamefont {de~la
  Ossa}(1990)}]{Candelas:1989js}%
  \BibitemOpen
  \bibfield  {author} {\bibinfo {author} {\bibfnamefont {P.}~\bibnamefont
  {Candelas}}\ and\ \bibinfo {author} {\bibfnamefont {X.~C.}\ \bibnamefont
  {de~la Ossa}},\ }\Doi {10.1016/0550-3213(90)90577-Z} {\bibfield  {journal}
  {\bibinfo  {journal} {Nucl. Phys.},\ }\textbf {\bibinfo {volume} {B342}},\
  \bibinfo {pages} {246} (\bibinfo {year} {1990})}\BibitemShut {NoStop}%
\bibitem [{\citenamefont {Kim}(2004)}]{Kim:2003vn}%
  \BibitemOpen
  \bibfield  {author} {\bibinfo {author} {\bibfnamefont {N.}~\bibnamefont
  {Kim}},\ }\Doi {10.1103/PhysRevD.69.126002} {\bibfield  {journal} {\bibinfo
  {journal} {Phys. Rev.},\ }\textbf {\bibinfo {volume} {D69}},\ \bibinfo
  {pages} {126002} (\bibinfo {year} {2004})},\ \Eprint
  {http://arxiv.org/abs/hep-th/0312113} {arXiv:hep-th/0312113} \BibitemShut
  {NoStop}%
\bibitem [{\citenamefont {Benvenuti}\ and\ \citenamefont
  {Tonni}(2009)}]{Benvenuti:2008bd}%
  \BibitemOpen
  \bibfield  {author} {\bibinfo {author} {\bibfnamefont {S.}~\bibnamefont
  {Benvenuti}}\ and\ \bibinfo {author} {\bibfnamefont {E.}~\bibnamefont
  {Tonni}},\ }\Doi {10.1088/1126-6708/2009/02/041} {\bibfield  {journal}
  {\bibinfo  {journal} {JHEP},\ }\textbf {\bibinfo {volume} {02}},\ \bibinfo
  {pages} {041} (\bibinfo {year} {2009})},\ \Eprint
  {http://arxiv.org/abs/0811.0145} {arXiv:0811.0145 [hep-th]} \BibitemShut
  {NoStop}%
\bibitem [{\citenamefont {Beisert}\ \emph {et~al.}(2003)\citenamefont
  {Beisert}, \citenamefont {Frolov}, \citenamefont {Staudacher},\ and\
  \citenamefont {Tseytlin}}]{Beisert:2003ea}%
  \BibitemOpen
  \bibfield  {author} {\bibinfo {author} {\bibfnamefont {N.}~\bibnamefont
  {Beisert}}, \bibinfo {author} {\bibfnamefont {S.}~\bibnamefont {Frolov}},
  \bibinfo {author} {\bibfnamefont {M.}~\bibnamefont {Staudacher}}, \ and\
  \bibinfo {author} {\bibfnamefont {A.~A.}\ \bibnamefont {Tseytlin}},\
  }\href@noop {} {\bibfield  {journal} {\bibinfo  {journal} {JHEP},\ }\textbf
  {\bibinfo {volume} {10}},\ \bibinfo {pages} {037} (\bibinfo {year} {2003})},\
  \Eprint {http://arxiv.org/abs/hep-th/0308117} {arXiv:hep-th/0308117}
  \BibitemShut {NoStop}%
\end{thebibliography}%

\end{document}